\documentclass[aps,superscriptaddress,twocolumn,floatfix,letterpaper,nofootinbib]{revtex4-1}

\usepackage{nicefrac}
\usepackage{braket}
\usepackage{amsmath, amssymb, graphicx}

\newcommand\toZ[1]{\lfloor #1 \rceil}
\newcommand\dd{\text{d}}
\newcommand\ii{\text{i}}
\newcommand\ee{\text{e}}

\newcommand\cM{\mathcal M}
\newcommand\cB{\mathcal B}

\newcommand\Z{\mathbb Z}
\renewcommand\th{\theta}

\renewcommand\th{\theta}
\renewcommand\t{\tilde}

\newcommand{\cN}{\mathcal{N}}

\newcommand\beq{\begin{equation}}
\newcommand\eeq{\end{equation}}

\usepackage[dvipsnames]{xcolor}

\begin{document}

\begin{titlepage}

\title{
A Lattice Chiral Boson Theory in $1+1$d
}

\author{Michael DeMarco}
\affiliation{Department of Physics, Massachusetts Institute of
Technology, Cambridge, MA, 02139, USA}
\affiliation{Co-Design Center for Quantum Advantage, Brookhaven National Laboratory, Upton, NY, 11973}
\email{mdemarco@bnl.gov}

\author{Ethan Lake}
\affiliation{Department of Physics, Massachusetts Institute of
Technology, Cambridge, MA, 02139, USA}

\author{Xiao-Gang Wen}
\affiliation{Department of Physics, Massachusetts Institute of
Technology, Cambridge, MA, 02139, USA}

\begin{abstract} 
Chiral field theories describe large classes of matter, from the edges of Quantum Hall systems to the electroweak sector of the Standard Model, but defining them on the lattice has been an ongoing challenge due to a no-go theorem precluding free local models, the potential of symmetry anomalies, and sign problems. Some approaches define a $1+1$d chiral field theory as the edge of a $2+1$d system and argue that the edge decouples from the bulk, but this can be difficult to verify due to finite size effects and strong interactions. On the other hand, recent work has shown how to define the $2+1$d bulk theory as an exactly solvable model with zero correlation length, in which case the edge theory may be extracted exactly.  
We use these techniques to derive a lattice field theory on a $1+1$d spacetime lattice which carries an anomalous chiral $U(1)$ symmetry with zero chiral central charge.  The lattice
theory with anomalous chiral $U(1)$ symmetry is always gapless, regardless of
lattice interactions.  
We demonstrate the chiral anomaly by coupling to a background
gauge field, develop a field theory which demonstrates the chiral behavior, and show how to assemble a chiral, anomaly-free theory where the gauge field may be taken to be dynamical.  
\end{abstract}


\maketitle

\end{titlepage}


\noindent

Between the fact that 
weak interaction couples left-hand and right-hand fermions differently and the
appearance of some of the most striking quantum anomalies in chiral models,
chiral quantum field theories (QFTs) have been the subject of enormous
interest. 
An essential tool in the study of quantum
field theories has been to regularize them on a lattice and to simulate their
behavior \cite{PhysRevD.25.2649}. Lattice QFT has proven enormously successful
in providing insight into the non-perturbative dynamics of quantum field
theories.  However, the lattice unfortunately does not mix easily with
chirality, and defining a chiral field theory on a lattice has remained a
challenge.

The first glimpse of the chiral problem was Nielsen and Ninomiya's theorem
\cite{NIELSEN1981219, NIELSEN1981173, NIELSEN198120}, which precludes the
appearance of non-interacting chiral fermions on the lattice for a large class
of theories. Instead, unwanted `doubling' modes appear, rendering the theory
non-chiral. A number of ingenious approaches have sidestepped this no-go theorem for
anomalous or anomaly-free chiral models \cite{Luscher:2000hn}, including the overlap-fermion
approach \cite{NARAYANAN199362, NARAYANAN1997360, Luscher:1998du,
Luscher:2000hn}, which computes correlation functions as the overlaps of
successive ground states \cite{PhysRevLett112247202}, and the related domain
wall approach \cite{KAPLAN1992342, SHAMIR199390, PhysRevD.65.074504}.  However, each of these comes with its
own drawback. In the domain wall approach, the applied gauge field propagates
in one higher dimension even for an anomaly-free theory, 
while the partition function in the overlap-fermion approach may not have an expression as a path
integral of local theory. 
Nonetheless, continued studies from both the lattice QFT and condensed matter communities then led to
a new class of theories similar to the domain wall theory. In this `mirror
fermion' approach \cite{Mirror1, Montvay:1992eg, Wen:2013ppa,Wang:2013yta,
PhysRevB.91.125147, Grabowska:2015qpk, Giedt:2007qg, PhysRevD.94.114504, 2021PhRvD.104a4503C, Wang:2018ugf, Zeng:2022grc}, the
$1+1$d lattice is understood as the edge of a $2+1$d manifold. The
chiral theory appears as a gapless theory on one edge, while its mirror
conjugate gapless theory resides on the other edge, with the bulk being gapped.
Taken together, the two gapless theories are non-chiral, but one seeks to
introduce interactions \cite{PhysRevB.81.134509, Tong:2021phe, PhysRevX.11.011063,Wang:2020iqc, Wang:2022ucy} that gap out only the mirror edge, which is always
possible for anomaly-free theory chiral theory (an insight from topological
order and symmetry protected topological (SPT) order in one higher
dimension \cite{W1313}). This approach introduces a compelling physical
picture, but comes with its own restrictions: it can only regularize
anomaly-free theories, and, more importantly, it relies on interactions to do
so and its validity is hard to confirm.

For both computational efficiency and insight into the underlying physics, we seek local lattice models of chiral QFTs that are regularized in the same dimension, whose chiral properties may be determined analytically, and that may be coupled to a gauge field. However, so far such theories have remained elusive. 

Recently, a whole new exact approach to gapped 2+1d $U(1)$ SPT phases on the
lattice has been found \cite{PhysRevLett.126.021603, DeMarco:2021erp}, and in
this paper we exploit these results to define a chiral boson theory in $1+1$d.
Our approach is most similar to the mirror-fermion approach above, but it
exhibits a critical feature: despite containing strong interactions, the 2+1d
bulk theory is exactly solvable with zero correlation length and so \emph{the edge may be
explicitly decoupled from the bulk}.  This exact solubility of the bulk model leads
to an 1+1d edge theory, which is local, well-defined, and contains a $U(1)$ 't
Hooft anomaly that can be seen analytically. This makes it easy to verify that
the theory has the correct chiral behavior, and makes our model of considerable
use for the study and simulation of chiral QFTs.  Because of the $U(1)$ 't
Hooft anomaly, we conjecture that the our 1+1d model is always gapless
regardless of lattice interactions, as long as the $U(1)$ symmetry is not
explicitly broken on lattice.  It is quite striking to see a lattice model
which remains gapless for any $U(1)$-symmetry preserving local interaction.

The $1+1$d chiral boson theory we present also paves the way for simulation of
more complicated chiral QFTs, including in higher dimensions and for nonabelian
symmetries. The simplest extension of this theory is a chiral fermion theory
that may be obtained by introducing a spin structure. Most importantly, we hope
that this result will spur continued collaboration between the condensed matter
and lattice QFT communities on these fascinating theories.

The extremely useful properties of the theory we study here follow from the
fact that it is a fixed-point theory. The key to writing a fixed-point theory
on the lattice was first discovered in \cite{Chen:2011pg} but not implemented
until recently \cite{PhysRevLett.126.021603, DeMarco:2021erp}: in order to
write down the topological action which produces a fixed-point theory, we must allow discontinuous
functions of the field variables. In turn, once we have a fixed-point
topological action in $2+1$d, we have a gapped bulk and a gapless $1+1$d edge
which decouple exactly, since the penetration of the gapless mode into the bulk
must be zero at the fixed point. This is what enables us to write down the pure
local $1+1d$ model and study its properties.

That we encounter physical quantities that are not
continuous functions of the field variables in fixed-point theories follows
from a simple argument. Consider, as we will shortly, a lattice QFT consisting
of $U(1)$ variables on the sites of a lattice. The space of field
configurations is $U(1)^{\text{n}_{\text{sites}}}$. We will need a function $\rho_v$ which indicates the vortex number on each plaquette. In a fixed-point theory, the output of $\rho_v$ should be an
integer for each plaquette, i.e. an element of $\mathbb
Z^{\text{n}_{\text{plaquettes}}}$. As a function from a connected manifold,
$U(1)^{\text{n}_{\text{sites}}}$, to a discrete space, $\mathbb
Z^{\text{n}_{\text{plaquettes}}}$, $\rho_v$ must be either discontinuous or
constant, and constant would be useless. Hence when describing vortices in a
fixed point theory, we should allow for discontinuous physical quantities; this
holds for topological defects in many theories.

Now let us turn to the model. We consider a spacetime lattice with sites
labeled by $i$ and a $U(1)$ variable $\phi_i$ on each site. To save many
factors of $2\pi$, we work with the $\phi_i$ quantized to unity, not $2\pi$, so
that all functions of $\phi_i$ must be invariant under $\phi_i\to \phi_i +
n_i$, with $n_i\in\mathbb Z$. We implement this as a gauge ``rotor
redundancy'':
\begin{equation}
    \phi_i \to \phi_i + n_i\label{eq:gaugeredphi}
\end{equation}
and will ensure that all physical quantities are invariant and the path
integral measure is gauge-fixed. 

The simplest way to define the theories we describe is to imbue all lattices
with a branching structure and use the differential $d$ and cup product $\cup$
from algebraic topology. We will not review the details of the formalism here
(see the supplemental material of \cite{PhysRevLett.126.021603} for full
details); instead we need only the following: A field which assigns a variable
to lattice sites, like $\phi_i$, is a 0-cochain. A field which assigns a
variable to all the $m$-dimensional cells of the lattice, an $m$-cochain. An
action must assign a real number to the three-dimensional cells of our lattice
and so is a 3-cochain. Our task is to construct a 3-cochain from the 0-cochain
$\phi_i$. We use the cup product, which takes an $m$-cochain $a_m$ and an
$n$-cochain $b_n$ to an $m+n$ cochain $a_m \cup b_n$ (we will abbreviate
$a_m\cup b_n$ as $a_mb_n$), and the lattice differential $d$, which takes an
$m$-cochain $a_m$ to an $m+1$ cochain $da_{m+1}$ and satisfies $\dd^2=0$. On the
lattice, $(\dd\phi)_{ij} = \phi_i - \phi_j$. Furthermore, we will
write out the most important equations, including the definition of the chiral
$1+1$d model, with explicit lattice indices.

\begin{figure}
    \centering
    \includegraphics[width = .3\columnwidth]{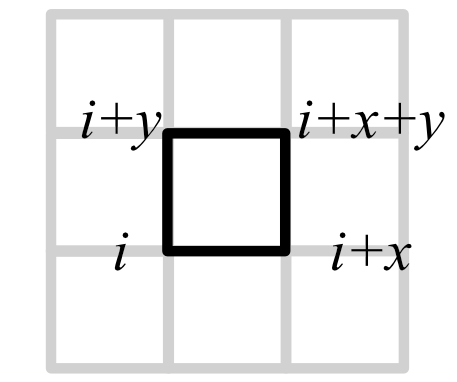}
    \caption{The model we study works on any lattice with a branching structure, but we will write down explicit expressions on this square lattice.}
    \label{fig:plaquette}
\end{figure}

We now return to the definition of a vortex referenced earlier. A branch cut in the field $\phi_i$ on the link $\braket{i, j}$ is labeled by:
\begin{equation}
    b_{ij} = \toZ{(\dd\phi)_{ij}} = \toZ{\phi_i - \phi_j}
\end{equation}
where $\toZ{x}$ denotes the nearest integer to $x$. Here, $b$ takes a non-zero value only on the links which cross a branch cut. Now, $\dd b = \dd\toZ{\dd\phi_{ij}}$ takes non-zero values only on the end of a branch cut, i.e. a vortex. Including a minus sign by convention, we define the vortex density in two dimensions as:
\begin{multline}
    \rho_v = -\dd\toZ{\dd\phi} = -\Big(\toZ{\phi_i - \phi_{i+x}} + \toZ{\phi_{i+x} - \phi_{i+x+y}} \\ - \toZ{\phi_{i+y} - \phi_{i+x+y}} - \toZ{\phi_i - \phi_{i+y}}\Big)\label{eq:vortexdensity}
\end{multline}
where we have written the term out explicitly on a square plaquette (see Figure \ref{fig:plaquette} and Appendix \ref{sec:Square}). 
One can think of $\rho_v$ as counting the branch cuts around the plaquette; it will be non-zero if a branch cut ends within the plaquette, which is when there is a vortex in the plaquette. Note that, while $b$ is not invariant under (\ref{eq:gaugeredphi}), $\rho_v$ is, as $\rho_v \to -\dd\toZ{\dd\phi + \dd n} = \rho_v - \dd^2n = \rho_v$. In three dimensions, we can define a vortex current in the same way:
\begin{equation}
    \jmath_v = \star(-\dd\toZ{\dd\phi})\label{eq:vortexcurrent}
\end{equation}
where we have introduced the lattice Hodge star operator. $\jmath_v$ is also invariant under $(\ref{eq:gaugeredphi})$.

The starting point of our 1+1d  model is a 2+1d fixed-point model with Hall conductance of $2k$, $k\in \mathbb Z$, derived in \cite{DeMarco:2021erp}. Let $\cN^3$ be a three-dimensional space-time lattice. The action is:
\begin{multline}
    S_k[\phi] = -2\pi \ii k \int_{\cN^3} (\dd\phi - \toZ{\dd\phi}) \cup \dd(\dd\phi - \toZ{\dd\phi}) 
    \\
    = 2\pi \ii k \int_{\cN^3} \dd\phi \cup \dd\toZ{\dd\phi} = - 2\pi \ii k \int_{\cN^3} \dd\phi \cup \rho_v\label{eq:3dAction}
\end{multline}
where used the fact that $\dd^2 = 0$, and that the action appears as $e^{iS}$ to simplify the action. Here ``$\int_{\cN^3}$'' means evaluation against a generator of the top cohomology of the lattice $\cN^3$. The full path integral is:
\begin{equation}
    Z = \int D\phi e^{\ii S_{k}[\phi]} \hspace{1 cm} \int D\phi = \prod_i \int_{-\frac{1}{2}}^{\frac{1}{2}} \dd\phi_i 
\end{equation}
where the integral measure is gauge-fixed under (\ref{eq:gaugeredphi}).

The most important aspect of the action (\ref{eq:3dAction}) is that it is a
total derivative, i.e. a surface term. It vanishes (mod $2\pi \ii$) on a closed
manifold. Hence we may evaluate it on a manifold $\cN^3$ with boundary $\cM^2 =
\partial \cN^3$ to obtain a theory solely on $\cM^2$:
\begin{equation}
 S_k =  \int_{\cM^2} 2\pi \ii k\phi \dd\toZ{\dd\phi} -L(\phi) 
=  \int -2\pi \ii k\phi \rho_v -L(\phi) \label{eq:2dAction}
\end{equation}
where we have added a possible additional non-topological term $L(\phi)$.  This is the
model which we wish to present and, together with its gauged version we will
see later, is the \emph{main result of this paper.} Writing out the indices for a square
lattice, the action is:
\begin{multline}
    S_k[\phi] = \sum_{i\in \cM^2}
-2\pi \ii k  \phi_i \Big(\toZ{\phi_i - \phi_{i+x}} + \toZ{\phi_{i+x} - \phi_{i+x+y}} \\ - \toZ{\phi_{i+y} - \phi_{i+x+y}} - \toZ{\phi_i - \phi_{i+y}}\Big) -L(\phi_i)
\end{multline}
where $i$ sums over the sites of the lattice.

Let us understand the properties of (\ref{eq:2dAction}) when $L(\phi)=0$. Under
the redundancy (\ref{eq:gaugeredphi}), the action is invariant, since $\rho_v$
is invariant and $n\rho_v$ is integer-valued and so does not affect the
exponential. The action also has particle-hole symmetry $\phi \to -\phi$ and
inherits the translation and rotation symmetry that the underlying lattice has. 

The term $2\pi \ii k\phi \dd\toZ{\dd\phi}$ has an unusual
$U(1)$ symmetry (Any additional terms $L(\phi)$ should have the usual symmetry $L(\phi+\theta) = L(\phi)$). Under $\phi\to\phi+ \theta$, with $\theta$ a global constant,
the action transforms as:
\begin{equation}
    S_k[\phi] \to S_k[\phi] - 2\pi \ii k \theta \int_{\cM^2} \rho_v
\end{equation}
If $\cM^2$ is closed, then the total vortex number is zero, as
$\int_{\cM^2}\rho_v = - \int_{\cM^2} \dd\toZ{\dd\phi} =0$ by summation by
parts, and so the theory is $U(1)$ symmetric. If $\cM^2$ has a boundary $\cB^1
= \partial \cM^2$, then the action is not $U(1)$ invariant.  Due to this unusual (anomalous)
$U(1)$ symmetry, adding any local $L(\phi)$ term cannot cancel the
$2\pi \ii k\phi \dd\toZ{\dd\phi}$ term.

The breaking of $U(1)$ symmetry in the presence of a boundary is our first
glimpse of the $U(1)$ anomaly. This perspective on anomalies, as symmetries
that break on the boundary, is more familiar to the condensed matter community.
It should be considered as a consequence of the familiar expression of the 't
Hooft anomaly, where the anomalous symmetry cannot be gauged, (which we
investigate next), as a background gauge field could impose an electric
potential that would create a boundary. 

Let us gauge the theory by coupling it to a background gauge field to
see the anomaly in a more usual light. At a first glance, it is not clear how
to gauge the action (\ref{eq:2dAction}), as it contains a term $\phi$ without a
derivative. We could proceed by integrating by parts to rewrite the action in
terms of ``$\dd\phi \toZ{\dd\phi}$,'' but this term is not manifestly invariant
under (\ref{eq:gaugeredphi}) and gauging it will break invariance under
(\ref{eq:gaugeredphi}). The obstruction of gauging is a common indication of a 't
Hooft anomaly.

One way to avoid this obstruction is to return to the 2+1d model
(\ref{eq:3dAction}).  Gauging such a model has been done in
\cite{DeMarco:2021erp} (see also Appendix C of \cite{DeMarco:2021dii}). The resulting action is:
\begin{multline}
    S_k[\phi;A ] = -2\pi \ii k \int_{\cN^3}\Big\{ (\dd\phi - A)(\dd A - \toZ{\dd A}) \\ - \toZ{\dd A} (\dd\phi - A)
    - \dd\Big[(\dd\phi - A)(\dd\phi - A - \toZ{\dd\phi - A})\Big]\Big\} \label{eq:gauged3dAction}
\end{multline}
Note that we also take $A$ to be periodic modulo unity. We also assume $A$ to be weak, in the sense that $\dd A - \toZ{\dd A} \approx 0$. This implies that $\dd\toZ{\dd A} = 0$, i.e. that field configurations are free of monopoles.

This action has three gauge redundancies. Two are of the same ``rotor redundancy'' form as before:
\begin{align}
    \phi_i\to \phi_i+n_i \label{eq:gaugeredunphi2}\\
    A_{ij} \to A_{ij} + m_{ij } \label{eq:gaugeredunA2}
\end{align}
for $n_{i}, m_{ij} \in \mathbb Z$. The third is the typical gauge invariance:
\begin{equation}
    \phi \to \phi + \theta \hspace{1 cm}
    A \to A + \dd\theta \label{eq:gaugereduntheta}
\end{equation}
for $\theta$ an $\mathbb R/\mathbb Z$-valued field. 

Now we separate (\ref{eq:gauged3dAction}) into boundary terms and bulk terms. We can rewrite the action as:
\begin{multline}
    2\pi \ii k \int_{\cN^3}
    \Big\{
    -A(\dd A - \toZ{\dd A}) + \toZ{\dd A}{A} - \dd(A(A-\toZ{A}))
    \Big\}
    \\ + 2\pi \ii k \int_{\cM^2} \Big\{
    \phi(\dd A - \toZ{\dd A}) - \toZ{\dd A}\phi 
    \\- (\dd\phi - A)(\dd\phi - A -\toZ{\dd\phi - A}) + A(A -\toZ{A})
    \Big\}
    \label{eq:separeted3dgaugedAction}
\end{multline}
We have now split the action into `bulk terms' consisting only of $A$ and `boundary terms' which contain all the $\phi$. Each of them are separately invariant under $\phi \to \phi+n$, and we have added and subtracted terms to ensure that they are invariant under $A\to A+m$. They are not separately invariant under the gauge symmetry (\ref{eq:gaugereduntheta}).

The boundary integral is the proper gauged action for the edge mode, i.e. our $1+1$d chiral boson model. Specifically, it is:
\begin{multline}
    S = 2\pi \ii k \int_{\cM^2} \Big\{
    \phi(\dd A - \toZ{\dd A}) - \toZ{\dd A}\phi 
    \\- (\dd\phi - A)(\dd\phi - A -\toZ{\dd\phi - A}) + A(A -\toZ{A})
    \Big\} \label{eq:2dActionGauged}
\end{multline}
This action is written on a square lattice in Appendix \ref{sec:GaugedActionSquare}.
Together with the ungauged $(A=0)$ model (\ref{eq:2dAction}), eq. (\ref{eq:2dActionGauged}) is the main result of this paper. One can check that the action is invariant under both of the symmetries (\ref{eq:gaugeredunphi2}) and (\ref{eq:gaugeredunA2}). However, it is not invariant under (\ref{eq:gaugereduntheta}), i.e. $\phi \to \phi + \theta$, $A\to A + \dd\theta$. The anomaly structure is in general complicated, but takes a simple form when we set $\dd\theta=0$. In that case, the action changes by a term:
\begin{equation}
    -2\pi (2k) i \theta \int_{M^2} \toZ{\dd A} \label{eq:anomaly}
\end{equation}
Now, $\int_{\cM^2}(\dd A - \toZ{\dd A})=-\int_{\cM^2}\toZ{\dd A}$ is the total flux of the gauge field over $\cM^2$, and so this is precisely $2\pi \ii (2k) \int \theta F$, i.e. the anomaly required by the Hall conductance. 
We should recognize this as the expected Adler-Bell-Jackiw anomaly, i.e. as a lattice, discrete generalization of $4\pi \ii k \int \theta \dd A$. As is usual, this failure of gauge invariance can be cancelled by an equal and opposite contribution from the bulk terms in eq. (\ref{eq:separeted3dgaugedAction}).

We have now examined the $1+1$d lattice model in detail and seen the $U(1)$ anomaly through both symmetry breaking in the presence of a boundary and through direct coupling to a background gauge field. Now we write down a continuum model for the edge theory and explain how it creates a chiral representation of $U(1)$ and a nonzero quantized Hall conductance. 

The first step is to to see that the topological term imparts a charge to vortices. We do this by examining the parent $2+1$d theory. Because vortices are proliferated in the model (\ref{eq:gauged3dAction}) and therefore are not well-defined excitations, we first confine them be adding in a term $\frac{1}{g}\sum_{\braket{i, j}}\cos 2\pi (\dd\phi - A)$. This term sets the vortices to be nearly zero, i.e. $\toZ{\star j} = 0$, in which case the action (\ref{eq:gauged3dAction}) can be written as a minimally coupled form of (\ref{eq:3dAction}):
\begin{equation}
    S = 2\pi \ii k \int_{\cN^3} (\dd\phi - A -\toZ{\dd\phi - A})\dd(\dd\phi - A - \toZ{\dd\phi - A})\label{eq:minimallyCoupled3d}
\end{equation} where we have ignored 1-cup products that encode framing (See \cite{DeMarco:2021dii} Appendix C). Coupling in the gauge field modifies the vortex current (\ref{eq:vortexcurrent}) to:
\begin{equation}
    \star \jmath = -\dd(\dd\phi - A -\toZ{\dd\phi-A})
\end{equation}
On a closed manifold $N^3$, the gauged bulk action (\ref{eq:minimallyCoupled3d}) can be rewritten as:
\begin{equation}
    2\pi \ii k \int_{N^3}(A\dd A + A\star \jmath + \jmath \star A)
\end{equation}
Hence the topological term leads to a charge $2k$ vortex. 

The ungauged action (\ref{eq:2dAction}) describes a bosonic field $\phi$ coupled to its vortices. To develop a continuum description, we recall the usual description of a compact field $\phi$ with its vortex field $\theta$:
\begin{equation}
    S \sim 2\pi \ii \int \left[\phi \partial_x \partial_t \theta + \frac{v}{2}(\partial_x \phi)^2 + \frac{v}{2}(\partial_x \theta)^2\right]
\end{equation}
Here $e^{2\pi \ii \hat \theta}$ creates a vortex in $e^{2\pi \ii \hat \phi}$, as can be seen from the commutation relations $[\hat\phi(x), \partial_{x'}\hat \theta(x')] = \frac{i}{2\pi} \delta(x - x')$. We have included velocity terms with speed $v>0$ that may be induced by a term $\sum_{\text{links}}\cos 2\pi \dd\phi$ in the lattice model, and have set $v_\theta = v_\phi =v$ for convenience. From the lattice model, we know that $\phi$ has charge $1$ and the vortex field $\theta$ has charge $2k$. We define the composite fields $\phi_R = \frac{1}{2}(\phi + \theta)$, $\phi_L = \frac{1}{2}(\phi - \theta)$ to get:
\begin{equation}
    2\pi \ii k \int(\phi_R\partial_x \partial_t \phi_R - \phi_L \partial_x \partial_t \phi_L + v (\partial_x \phi_R)^2 + v(\partial_x \phi_L)^2)
\end{equation}
Thus the chiral model consists of a right-moving mode $\phi_R$ and a left moving mode $\phi_L$, which have respective equations of motion $(\partial_x \pm \partial_t)\partial_x\phi_{L/R} = 0$, where $\rho_{L/R} = \partial_x \phi_{L/R}$ is the usual bosonized excitation density. As there are equal numbers of left and right moving modes, there is no gravitational anomaly or, equivalently, thermal Hall conductance.

On the other hand, there is a $U(1)$ anomaly which arises because $\phi_L$ and $\phi_R$ have differing charges. Denote the change of a field $\varphi$ by $C[\varphi]$, so that $C[\phi] = 1$. We have seen that vortices have charge $2k$ and so $C[\theta] = 2k$. Hence the anomaly has coefficient:
\begin{equation}
    C[\phi_R] - C[\phi_L] = C[\theta] = 2k
\end{equation}
This is consistent with the Hall conductance of the bulk system, which is $2k\frac{e^2}{h}$ \cite{DeMarco:2021erp}.

We have seen how to create a $1+1$d lattice theory which realizes an anomalous chiral gapless field theory with a background gauge field. In order to make the gauge field dynamical, we would like to create an anomaly-free, but chiral gapless field theory. The solution is to layer multiple copies (say $N$) of the system with differing levels $k_I$ and charges $q_I$, $I=1, ... N$. Each layer contributes an anomaly factor of $k_Iq_I^2\mathcal A$, where $\mathcal A$ is the anomaly factor defined in Appendix \ref{sec:AnomalyApp}. This leads to a familiar anomaly cancellation condition:
\begin{equation}\label{eq:AnomalyCancellation}
    \sum_{I=1}^N k_I q_I^2=0
\end{equation}
If $k_I$ and $q_I $ are chosen to satisfy this, then an anomaly-free chiral lattice field theory is:
\begin{multline}
    S = 2\pi i\sum_{I=1}^N k_I \int_{\cM^2} \Big\{
    \phi_I q_I (dA - \toZ{dA}) - q_I\toZ{dA}\phi_I 
    \\- (d\phi_I - q_I A)(d\phi_I - q_IA -\toZ{d\phi_I - q_IA})
    \Big\} \label{eq:2dActionGauged_AnomFree2}
\end{multline}
and $A$ may be interpreted as a dynamical gauge field. The fields with $k_I>0$ are 'right-moving,' while those with $k_I<0$ are left moving; accordingly these theories carry chiral $U(1)$ representations, for example with $(k_I, q_I) = \{(1, 3), (1, 4), (-1, 5)\}$. A generalized anomaly construction with manifestly on-site $U(1)$ symmetry is given in Appendix \ref{app:AnomalyOnSite}.

We have written down a boson theory with chiral $U(1)$ symmetry in $1+1$d by extracting the edge theory from an exactly soluble $2+1$d chiral model. We then demonstrated the $U(1)$ anomaly by both inspection of the ungauged theory and by explicitly coupling in a background gauge field and calculating the variation of the edge action. Finally, we wrote down a continuum theory for our model and showed that it carries chiral $U(1)$ charge as expected. The key to our model is the expression for the vortex density \eqref{eq:vortexdensity} which allows for a fixed-point description of the topological defects of the field. The generalization to other models with more complicated target spaces (e.g. $SO(3)$) will make use of similar discontinuous functions. 

\emph{Acknowledgments} This research was partially supported by NSF DMR-2022428, the NSF Graduate Research Fellowship under Grant No. 1745302, by the Simons Collaboration on Ultra-Quantum Matter, which is a grant from the Simons Foundation (651440), and by the U.S. Department of Energy, Office of Science, National Quantum Information Science Research Centers, Co-design Center for Quantum Advantage (C2QA) under contract number DE-SC0012704. MD and EL acknowledge useful discussions with H. Goldman,  J.Y. Chen, J. Wang, and J. Wen. MD acknowledges useful discussions with V.V. Albert, and is grateful to E. Witten for comments on generalizing to $SU(2)$. 

\appendix

\bibliography{ChiralBoson}

\section{Adding Dynamics to the Chiral Theory}
In the form presented, chiral theories in $1+1$d typically do not contain dynamical information in the topological term. For a typical continuum Lagrangian:
\begin{equation}
    2\pi [i m \partial_x \phi \partial_t \phi + v(\partial_x\phi)^2]
\end{equation}
whether the model is a left-mover or a right-mover is encoded in the sign of the coefficient $m$ of the topological term. On the other hand, the velocity of this moving mode is given by the non-universal coefficient $v$. In the Hamiltonian picture, $v(\partial_x \phi)^2$ is the entire Hamiltonian; the topological term instead modifies the commutation relations.

Similarly, the lattice model we presented in this paper contains trivial dynamics. To ensure that the left- and right-moving modes in our model actually move, we add a non universal term to the action, namely:
\begin{equation}
    S_{k, v}\equiv 2\pi \ii k \int \phi \dd\toZ{\dd\phi} + 2\pi v\sum_{\text{links}}(\dd\phi - \toZ{\dd\phi})^2
\end{equation}
Here $v$ plays the role of the velocity in the continuum. It is tempting to think that as $v\to \infty$, the topological term could be dropped, as $\dd\phi -\toZ{\dd\phi} \approx 0$ implies that $\dd\toZ{\dd\phi} = 0 $, and then the effective field theory is $v[(\partial_x\phi)^2 + (\partial_t\phi)^2]$. However, this is not true: the topological term modifies the commutation relations in the Hamiltonian picture, and modifies the response to a background gauge field, even when $v\to \infty$.



\section{Full Anomaly}\label{sec:AnomalyApp}
Under a gauge transformation, the action \ref{eq:2dActionGauged} at level $k$ changes by a factor of $k\mathcal A$, where:
\begin{multline}
    \mathcal A \equiv S_{k=1}[\phi+\theta; A-d\theta] - S_{k=1}[\phi; A]
    \\=
        2\pi i \int_{\cM^2} \Big\{
        \theta(dA - \toZ{dA}) - \toZ{dA}\theta
        \\+ d\theta A + A d\theta + d\theta d\theta
        + A\toZ{A} - (A+d\theta)\toZ{A+d\theta}
    \Big\}
\end{multline}
Taking $d\theta = 0$ leads to eq. \eqref{eq:anomaly}. 

\section{Cup Products and the Lattice Differential on Square and Cubic Lattices}\label{sec:Square}
Typically, the cup product and lattice differential are defined on simplicial complexes. However, they may be generalized to square and (hyper-) cubic lattices. Mathematically, the differential $d$ is induced by the boundary operation, while the cup product $\cup$ can be derived from the action on cohomology. For our purposes, the simplest way is to add links so that the lattice becomes simplicial, and then to set all fields on those links to zero.

\begin{figure}
    \centering
    \includegraphics[width = .3\columnwidth]{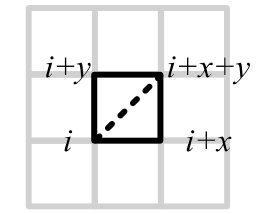}
    \caption{We may extend the square lattice into a triangular lattice by adding links (dotted). We can return to the square lattice, while retaining the cup product and lattice differential from the triangular lattice, by setting all fields on the dotted links to zero.}
    \label{fig:plaquetteTriangle}
\end{figure} 

For example, let $\varphi^\Delta$ be a zero-cochain and $c^\Delta$ be a two-cochain on a two-dimensional triangular spacetime lattice, such as that shown in Figure \ref{fig:plaquetteTriangle}. The sum $\int \varphi^\Delta \cup c^\Delta$ would contain two terms with $\varphi^\Delta_i$:
\begin{equation}
\varphi^\Delta_i c_{ijl}^\Delta - \varphi^\Delta c_{ikl}^\Delta
\end{equation}
We can then group these into square-lattice variables $c_{ijkl}^\Box = c_{ijl}^\Delta - c_{jkl}^\Delta$ and $\varphi_i^\Box = \varphi_i^\Delta$, and so the sum $\int \varphi^\Box \cup c^\Box$ contains a single term with $\varphi^\Box_i$:
\begin{equation}
\varphi^\Box_i c_{ijkl}^\Box
\end{equation}
For a one-cochain $a^\Delta$, we can use this idea to define the lattice differential:
\begin{equation}
(da^\Box)_{ijkl} = da^\Delta_{ijl} - da^\Delta_{ikl} = a_{ij} + a_{jl} - a_{ik} - a_{il}
\end{equation}
or the cup product:
\begin{equation}
(a^\Box\cup a^\Box)_{ijkl}
=
(a^\Delta\cup a^\Delta)_{ijl} - (a^\Delta\cup a^\Delta)_{ijl}
=
a_{ij}a_{jl} - a_{ik}a_{kl}
\end{equation}
These are the relations needed to construct eq. \eqref{eq:2dGaugedActionSquareLattice}.

\section{Gauged Action on a Square Lattice}\label{sec:GaugedActionSquare}
We may use the techniques of Appendix \ref{sec:Square} to write the gauged action \eqref{eq:2dActionGauged} on a square lattice as:
\begin{widetext}
\begin{multline}\label{eq:2dGaugedActionSquareLattice}
    S = 2\pi i k \sum_i \phi_i \Big(
    A_{i, i+x} + A_{i+x, i+x+y} - A_{i+x, i+x+y} -
    A_{i+y, i+x+y} - \toZ{A_{i, i+x} + A_{i+x, i+x+y} - A_{i+x, i+x+y} -
    A_{i+y, i+x+y}}\Big)
    \\ - \toZ{A_{i, i+x} + A_{i+x, i+x+y} - A_{i+x, i+x+y} -
    A_{i+y, i+x+y}} \phi_{i+x+y}
    \\
    - (\phi_{i} - \phi_{i+x} - A_{i, i+x})(\phi_{i+x} - \phi_{i+x+y} - A_{i+x, i+x+y} - \toZ{\phi_{i+x} - \phi_{i+x+y} - A_{i+x, i+x+y}})
    \\
    + (\phi_{i} - \phi_{i+y} - A_{i, i+x
    y})(\phi_{i+y} - \phi_{i+x+y} - A_{i+y, i+x+y} - \toZ{\phi_{i+y} - \phi_{i+x+y} - A_{i+y, i+x+y}})
    \\
    + A_{i, i+x}(A_{i+x, i+x+y} - \toZ{A_{i+x, i+x+y}}) - A_{i, i+y}(A_{i+y, i+x+y} - \toZ{A_{i+y, i+x+y}})
\end{multline}
\end{widetext}

\section{Anomaly-Free Theory with On-Site $U(1)$ Symmetry}\label{app:OnSite}

To create an anomaly-free chiral gapless lattice field theory with on-site symmetry, we consider several fields $\phi_I$, $I=1, ... N$, with the action:
\begin{align}
\label{SphiI}
 S = 2\pi \ii \int_{\cM^2} \sum_{I,J} k_{IJ} \phi_I \dd&\toZ{\dd\phi_J}\\
+ h_{IJ} \dd&\big( \phi_I(\toZ{\dd \phi_J} - \dd \toZ{\phi_J})\big)
,
\end{align}
where $k_{IJ},h_{ij} \in \Z$ and may not be symmetric.  The new action amplitude
$\ee^{-S}$ has a $\Z$-gauge invariance $\phi_I \to \phi_I+ n_I$.  It also has a
$U(1)$ symmetry $\phi_I\to \phi_I+ q_I \th$, where $q_I\in \Z$ and are coprime
to each other, even when $\cM^2$ has a boundary, provided that
\begin{align}
 q^\top k q = 0, \ \
 q^\top h = q^\top k
\end{align}
\begin{align}
v^\top  h = 0,~~~\forall~~v \in \{v: v^\top  q=0\}.
\end{align}

To simplify our discussion, we assume $\cM^2$ to have no boundary and drop the
total derivative term.  For the above choice of $k_{IJ}$, then an anomaly-free
chiral lattice field theory with $U(1)$ gauge field is described by the
following action amplitude $\ee^{-S}$:
\begin{align}
     \label{eq:2dActionGauged_AnomFree}
&  \ee^{ 2\pi \ii \sum_{I,J} k_{IJ} \int_{\cM^2} 
    q_J\phi_I (dA - \toZ{dA}) - q_I \toZ{dA}\phi_J}\times \nonumber
    \\
&  \ee^{ 2\pi \ii \sum_{I,J} k_{IJ} \int_{\cM^2} 
    (\dd \phi_I -q_I A)(\dd \phi_J -q_J A - \toZ{\dd \phi_J -q_J A})}\times
\\
&  \ee^{-\sum_{I,J} V_{IJ} \int_{\cM^2}
 (\dd \phi_I -q_I A - \toZ{\dd \phi_I -q_I A}) 
\star (\dd \phi_J -q_J A - \toZ{\dd \phi_J -q_J A})}
\nonumber 
\end{align}
and $A$ may be interpreted as a dynamical gauge field.  Here we have included a
possible dynamical `velocity' term $V_{ij}$.  The topological $k_{IJ}$ term breaks the
time-reversal and reflection symmetry. Such a symmetry breaking cannot be
continuously tuned to zero since $k_{IJ}$ are quantized as integers.  If there
is no integer invertible matrix $W$ to make $k \to W kW^\top =-k$, then the
theory breaks time-reversal and reflection symmetry in a topological way, and
is a chiral theory.

\section{The Anomaly Condition}\label{app:AnomalyOnSite}

We can make a field redefinition $\phi_I = W_{IJ}\t\phi_J$ where $W_{IJ}\in \Z$
and det$(W)=1$:
\begin{align}
 S &= 2\pi \ii \int_{\cM^2} \sum_{I,J} \t k_{IJ} \t \phi_I \dd\toZ{\dd\t\phi_J}
+ \t h_{IJ} \dd\big( \t\phi_I (\toZ{\dd \t\phi_J} - \dd \toZ{ \t\phi_J}\big)
\nonumber\\
\t k &= W k W^\top,\ \ \ \ \t h = W h W^\top.
\end{align}
The $U(1)$ transformation becomes
\begin{align}
 \t\phi_I \to \t \phi_I + \t q_I \th,\ \ \ \
\t q = W^{-1} q.
\end{align}
We can always find an invertible integer matrix $W$ to make $\t q
=(1,0,\cdots,0)$. To show this, we consider the following Smith normal form
of a $N\times N$ matrix
\begin{align}
U \begin{pmatrix}
 q_1 & 0 & \cdots & 0 \\
 q_2 & 0 & \cdots & 0 \\
 \vdots & \vdots & \ddots & \vdots \\
 q_N & 0 & \cdots & 0 \\
\end{pmatrix} V = 
\begin{pmatrix}
 1 & 0 & \cdots & 0 \\
 0 & 0 & \cdots & 0 \\
 \vdots & \vdots & \ddots & \vdots \\
 0 & 0 & \cdots & 0 \\
\end{pmatrix}
\end{align}
where we can always find invertible integer matrices $U$ and $V$ to satisfy the
above relation, since $q_I$ are coprime to each other.  The $W$ matrix that we
want is given by $W=U^{-1}$.

In terms of the new fields, we find
the $U(1)$ symmetry is anomaly-free if 
\begin{align}
\t k_{11}= \t q^\top \t k \t q = q^\top k q =0 .
\end{align}
Indeed, if we choose the total derivative term coefficients to be
\begin{align}
\t h_{11} =&~~0 \\
\t h_{1J}\big|_{J\geq 2}=& -\t k_{1J} \big|_{J\geq 2}
\\
\t h_{IJ}\big|_{I\geq 2} =&~~0 
\end{align}
we find that the action amplitude $\ee^{-S}$ to be invariant under both the
$U(1)$ transformation $\t \phi_1 \to \t \phi_1 +\th$ and the $\Z$-gauge
transformation $\t \phi_I \to \t\phi_I + n_I$ even when $\cM^2$ has boundaries.
This implies that the $U(1)$ symmetry is indeed anomaly-free.

From the known $\t h$, we can obtain $h =W^{-1} \t h (W^\top)^{-1}$.
The condition on $\t h$ can be written in terms of $h$:
\begin{align}
\t q^\top \t h &= \t q k, &
\t v^\top \t h &= 0,  \text{ for any }  \ \t v^\top \t q=0 .
\nonumber\\
 q^\top h &= q^\top k, & 
v^\top  h &= 0,  \text{ for any } \  v^\top  q=0.
\end{align}

Last, to obtain eq. \eqref{eq:anomaly}, we note that under a gauge
transformation, the action \eqref{eq:2dActionGauged} at level $k$ changes by a
factor of $k\mathcal A$, where:
\begin{multline}
    \mathcal A \equiv S_{k=1}[\phi+\theta; A-d\theta] - S_{k=1}[\phi; A]
    \\=
        2\pi i \int_{\cM^2} \Big\{
        \theta(dA - \toZ{dA}) - \toZ{dA}\theta
        \\+ d\theta A + A d\theta + d\theta d\theta
        + A\toZ{A} - (A+d\theta)\toZ{A+d\theta}
    \Big\}
\end{multline}
Taking $d\theta = 0$ leads to eq. \eqref{eq:anomaly}.

\end{document}